# ON COMPARISON OF SIMULATED AND OBSERVED SEISMICITY

Aleksandr M. Linkov[a,b], Liliana Rybarska-Rusinek[c*], Victor V. Zoubkov[d]

[a]*Institute for Problems of Mechanical Engineering of Russian Academy of Sciences (Saint Petersburg, Russia)*
[b]*Saint Petersburg State Polytechnic University (Saint Petersburg, Russia)*
[c]*Rzeszow University of Technology (Rzeszow, Poland)*
[d]*National Mineral Resources University (Saint Petersburg, Russia)*
[*]*Corresponding author: rybarska@prz.edu.pl, Rzeszow University of Technology, The Faculty of Mathematics and Applied Physics, al. Powstańców Warszawy 8, 35-959 Rzeszów, Poland*

**ABSTRACT**

Numerical simulation of seismicity has been successfully developed and used for the two last decades. Presently, the general theory of modeling and the progress in computational techniques provide wide options for simulation of seismic and aseismic events with various source mechanisms accounting for blocky structure of rock mass, inclusions, faults, cracks, complicated contact conditions and various mechanical properties of rock. Meanwhile, in practical applications, the input data are limited and uncertain. The data on observed seismicity are also often limited with a few parameters, like coordinates and time. The paper aims to agree the input and output data, used in and provided by numerical simulations, with uncertain and limited data of direct observations. For the input parameters, we suggest their minimal set, which complies with commonly available data. For output seismic parameters, we distinguish three major groups, which are provided by field observations. The first group includes the common (minimal) data on distributions of the event location. These distributions are of special value for improving the input data on geometrical features of a problem. The second group employs the data (commonly available, as well) on the event magnitude. These distributions are of exceptional need for evaluating the risk of strong events. The third group employs data on the event source mechanism. It is based on the tensor of seismic moment/potency, provided by advanced mining seismic systems. This group includes distributions of the geometrical parameters of the event source (orientation of nodal planes, B, P and T directions). It is especially important when establishing and using the connection between stresses and seismicity. The exposition is illustrated by considering an example of long-wall mining in a coal seam.

## 1. INTRODUCTION

Starting from the pioneering work by M. D. G. Salamon [1], numerical simulation of seismicity has been developed and used for mining applications (e.g. [2-10]). A comprehensive review is given in the paper [8], available for free on line. It also presents in detail the general theory of simulation seismic and aseismic events. Thus there is no need to dwell on these issues. Note only that the theory supported with the progress in computers and computational techniques allows one to simulate both mechanical and seismological quantities accounting for blocky structure of rock mass, faults, inclusions, cracks, complicated interface conditions, 3D geometry of excavations, various mechanical properties of rock and various mechanisms of events. The input and output data may be quite detailed. It appears that the wide options suggested by modeling do not conform to the limited and uncertain input parameters available in mining practice, while the simulated output data may be well beyond the limited number of quantities reliably measured by a mining seismic unit. Indeed, in many cases, the local geological features around an excavation are unknown; in-situ stresses, even when measured, are frequently inaccurate; mechanical properties of rock and interfaces are rather uncertain. Commonly, seismic observations in mines provide at most time and roughly estimated location and energy of a seismic event. Therefore in practical applications of joint numerical modeling of stress changes and accompanying seismicity, we need to agree input and output data of a model with limited and uncertain data of direct measurements and observations in practice. The paper aims to make a step in finding a balance, which corresponds to the recent progress in geomechanics and applied seismology.

Clearly, an optimal set of input and output parameters involved in calculations depends on a particular application of modeling. The latter may be employed for various purposes including (i) interpretation of observed seismicity, (ii) calibration and improvement of input parameters, (iii) assessment of a current state of rock mass, (iv) playing scenarios for making a choice between mining plans; (v) sensitivity analysis. The set also depends on a level of geomechanical and seismological service at a particular site. Still, for the objectives (i), (ii) and (iii), there are general features which may be combined in a set close to optimal when employing the advanced techniques of modern geomechanical and seismic observations.

The papers [7, 9, 10] containing applications of the general theory to case studies, may serve to compare the approaches employed to choose input and output parameters. Specifically, in contrast with [7], the recent papers [9, 10] employ a *smaller* number of *input parameters* and *greater* number of *output quantities*.

The *reduction of the input information* is reached by prescribing location, orientation and size of source-flaws in a way different from that used by Salamon [1], which consisted in seeding the flaws randomly. The authors check the Coulomb criterion for planes of various orientations at points of prescribed meshes. The isolated lobes, where the criterion is met, are identified as seismic sources with prescribed coordinates, orientation and sizes. The sources, that are located within the source radius of a larger source, are excluded. In the line of the general theory [8], the events corresponding to the initial geometry, are excluded from further consideration: only events occurring on mining steps are taken into account.

This approach to seeding the sources avoids employing *a priori* assumptions about the sizes, orientation and location of flaws. It becomes equivalent to that by Salamon as concerns with the location and orientation of the flaws when the density of statistically seeded flaws is high enough in the considered area. Then the difference in seeding appears mostly in sizes of the sources. In the statistical approach [1] they are distributed randomly with a prescribed mean value; the minimal size may be close to zero. In the approach of [10], the minimal size cannot be less than the mesh size. The limitation on the minimal size of flaws leads to the computational effect noted in [10]: "The modeled seismicity has much more middle-size and much less small-size events as compared to the observed seismicity". This may distort the frequency-magnitude dependence. We think that neglecting *a priori* information on the average size of sources is not necessarily an advantage, especially when taking into account that for registered events their average size depends on the used sensitivity of geophones. Note also that in computational sense there are no significant differences between the two ways as concerns with time and memory expense for seeding. Below, when presenting an example, we follow the path by Salamon of statistical seeding the flaws to employ data of seismic observations in a mine on average size of sources and to keep the option to compare simulated dependence frequency-magnitude with that following from observations.

A valuable *extension of the output quantities,* as compared with [7], is reached in [9, 10] by making use of the *advanced seismic techniques*, such as recovering the source mechanism from amplitudes and polarities of P- and S-waves [11] and stress inversion [12-14]. The authors use parameters of nodal planes to compare the observed and simulated seismicity. Plotting their stereographic projections provides illuminating pictures, which clearly reveal geometrical features of the sources.

In the present paper, we combine and compliment the sets of input and output parameters, used in the papers [7] and [10], to obtain a set, which presents a reasonable choice for case studies. It looks optimal in the sense that being not excessive, it agrees with the advanced level of mining seismology. Specifically, similar to papers [7, 9, 10], the number of *input parameters* is reduced to a reasonable minimum. The *output parameters* are allocated to three groups with quite distinct practical applications. The first one refers to *location* of events and includes spatial and temporal distributions. The second refers to distributions of scalar characteristics of event *strength*, like energy, magnitude of seismic moment or seismic potency. The third refers to distributions of parameters characterizing geometrical features of the event *mechanism*. To distinguish an optimal set of output data, we follow the line of the papers [9, 10] and compliment the output spatial and temporal distributions with distributions of parameters characterizing the source mechanism, and employ stereographic projections to visualize its geometrical features. Meanwhile, we preserve statistical seeding of flaws. This serves us firstly to account for the commonly available data on a particular level of average energy of events recorded. Secondly, this extends options for comparison of simulated and observed seismicity by using scalar characteristics of the event strength in dependencies of the "frequency-magnitude" type. This provides options to quantify the risk of strong events by using the results presented in [15, 16]. The exposition is illustrated by considering an example of long-wall mining in a coal seam.

## 2. INPUT DATA FOR A BASIC CODE AND SIMULATION OF EVENTS

*2.1. Input data on mining conditions*

Numerical simulations of seismicity are performed by complimenting a conventional basic computer code of BEM, FEM or DEM with a number of 'seismic' subroutines (see, e.g. [1-10]). The input data of a basic code include (i) initial geometry of mined area and geological features to be accounted for, (ii) changes of the geometry on mining steps, (iii) boundary conditions and conditions at interfaces of structural blocks, and (iv) in-situ stresses. A particular choice of input parameters depends on a used basic code, information available and parameters to be studied when playing various scenarios. In this work we employ as the basic code an updated version of the program FAULT-3D [17]; it is founded on the 3D hypersingular boundary element method. The basic code is complimented with subroutines, described in detail in [8] and serving for simulation of seismic and aseismic events.

Fig. 1  For illustration, we consider the input geometry, which corresponds to mining of the seam Fifth in the north part of the mine Severnaya of the Vorkuta coal district in April 2013 (Fig. 1). The depth of mining is $H = 870$ m. The seam is flat and about horizontal (the dip angle is $5^{\circ}$). The mined out area is dashed in Fig. 1. The panel to be extracted by the longwall 512-z_V is shown crisscrossed. The longwall is of the length $L_w$ = 300m, its strike line is approximately along the East direction, the stope moves in the South direction. During April 2013, the propagation was about 100 m; thus the stope advances approximately 3 m per day. We represent the corresponding changes of the mined area by 10 mining steps. Each of them adds a strip of 10 m width and 300 m length what simulates the wall propagation for roughly three days.

The mechanical properties of rock, boundary conditions and in-situ stresses are prescribed in accordance with available information as follows. The elasticity modulus of rock is $E$ = 37000 MPa, the Poisson's ratio is $v$ = 0.25, the principal vertical stress is $\sigma_{11}$ = -21.75 MPa. In view of large area of subsidence, the normal pressure of the roof on the floor is significant and it should be taken into account. We account for it by employing the observations and theory summarized in [18]. It is established that the in-situ normal pressure is recovered at the distance $H\cot\varphi$ from unmined edge of a seam, where $\varphi$ is so-called pressure-angle. For the considered mine, $\varphi$ = 70°. The subsidence of roof into the mined out area influences also the confining pressure decreasing it as compared with the magnitudes of horizontal in-situ stresses. The reasonable values of the confining coefficients along West-East and North-South lines become respectively, 0.7 and 0.5. Thus we accept $\sigma_{22}$ = 0.7 $\sigma_{11}$ = -15.225 MPa, $\sigma_{33}$ = 0.5 $\sigma_{11}$ = -10.875 MPa as the principal stresses in these directions.

This completes prescribing the input data used by the basic code. The data are actually the same as those routinely used for simulation of rock pressure in the Vorkuta coal district.

*2.1. Input data on initial flaws*

The input data on the possible sources of seismic events include the data on their (i) location and orientation, (ii) size, (iii) shear rigidity, (iv) tensile and shear strength, (v) shear softening modulus, (v) portion of aseismic events. For reasons explained in Introduction, in the present paper we use statistical seeding of initial flaws. The flaws are seeded in a sufficiently large volume around the stope. It is assumed to be a parallelepiped 100 m high, 400 m long in the strike direction (West-East) and 250 m wide along the propagation line (North-South). Its horizontal cross-section is located symmetrically about the seam middle plane. Its West side is located at the distance of 50 m to the West from the left edge of the mined out area shown in Fig. 1. Its North side is at the distance of 150 m behind the initial position of the stope on April 1, 2013. The volume of seeded flaws is $V$ = 100 x 400 x 250 = $10^7$ m$^3$.

The average length of flaws to be seeded is prescribed by using the data of seismic observations in the mine. They imply that the average energy $W_{av}$ of the events recorded during April was 1600 J. By [8], the corresponding average size of flaws is $l_{av} = 1.21\sqrt[3]{W_{av}E/c_{0\tau}^2}$, where $c_{0\tau}$ is the initial cohesion at a flaw surface. Assuming $c_{0\tau}$ = 2.5 MPa, this yields $l_{av}$ = 2.56 m. To have a representative number of events, the total number of seeded flaws is taken $N$ = 20000. Therefore, the average distance between the flaws is $L_{av} = \sqrt[3]{V/N}$ = 7.9 m. The corresponding density of flaws defined as $\xi = l_{av}/L_{av}$, is $\xi$ = 0.32 what is within the range 0.14 < $\xi$ < 0.75, recommended in [6, 8].

The location of flaw centers, their dip and strike angles are prescribed randomly with uniform distributions. The size $l$ of a flaw is prescribed randomly with the exponential distribution of probability density: $f(l) = \left[\exp\left(-\frac{l}{l_{av}}\right)\right]/l_{av}$. When having the size $l$, we obtain the shear rigidity of embedding rock as $K_S$ = 0.9$E/l$ [8].

The tensile strength of a flaw is assumed zero. The shear strength is prescribed by the Coulomb friction law with the mentioned initial cohesion $c_{0\tau}$ = 2.5 MPa and the friction angle $\rho$ = 15$^o$. The residual cohesion is zero.

In this study, having no data on aseismic deformation of rock, we do not account for aseismic (creeping) events. Thus the parameters defining the portion of aseismic events are set to have merely seismic events.

This completes prescribing the data on flaws. Seeding is performed by the subroutine *FlawInput* included into the basic code.

## 3. OUTPUT DATA AND THEIR ANALYSIS

In addition to common mechanical data on stresses, strains and displacements, provided by a conventional basic code, we obtain data on seismic events caused by changes of stresses at initial flaws. These data are similar to those for events observed in mines. Specifically we obtain temporal and spatial distributions of events. For each simulated seismic event, we have coordinates of its source, orientation of its plane, its size, vector of displacements, the stress drop and the type (tensile of shear). These data define the energy of an event, its potency tensor and seismic moment tensor (see, e. g. [16]). In particular, we obtain the data on the mechanism of an event, such as orientation of nodal planes, zero B direction, P and T directions. What is of special significance, in contrast with observed seismicity, for any event we exactly know the mechanical state at which the event occurred. The statistical distributions of properly chosen output parameters may provide illuminating pictures, which facilitate better comprehension of the rock state. The comparison with the observed events serves to improving the input data, to suitable interpretation of observed seismicity and, as a result, to making confident practical decisions. The questions are: which distributions are the best candidates for an analysis and how to use them?

We distinguish three groups of statistical distributions, involving, respectively, (i) locations, (ii) magnitudes, and (iii) geometrical characteristics of events. The first group gives understanding of geometrical and geological features. The second group provides estimation of the risk to have a strong dangerous event like rockburst. The third group clarifies the influence of mechanical factors such as in-situ stresses and strength parameters. We illustrate the choice of distributions of each of the three groups by considering the output results for the example of long-wall mining.

*3.1. Temporal and spatial distributions of event locations*

In this and following subsections we present the results of numerical simulation of seismicity with the input data of Section 2. The calculations are performed by using a modified version of the code SEISM-3D [6, 7].

In the problem considered, *temporal* changes are associated with the mining steps. As mentioned, a step of 10 m corresponds roughly to three days. Therefore, we can trace changes in time by comparing distributions on successive mining steps. From the calculations, it appears that the number of simulated events, their grouping about the moving longwall, energy and mechanism parameters do not change significantly from step to step. This can be expected because, as clear from Fig. 1, the geometry is actually reproduced when mining occurs in the middle part of a panel. Hence, in the considered case, it is sufficient to focus on a typical, say first, mining step.

Typical spatial distributions of events are shown in Fig. 2 and 3. Fig. 2 presents the projection of event locations to the seam plane. The solid lines show the opening edges at the beginning of the step; the dashed line corresponds to 10 meters advance of the stope. It can be seen that the events group along the mined strip. The picture agrees with the distribution of events observed in the mine and shown in Fig. 1. Therefore, the model used captures geometrical features of the case. Note however that during the considered period (April 2013) there were also registered seismic events not shown in Fig. 1 because they were well apart from the stope 412-3-V. These events, located at a distance of some 400 m, were grouped around another stope at a neighbor seam mined simultaneously. Thus, if wanting to account for all the events occurred in April, it would be necessary to complement the geometrical scheme with the geometry of openings and mining steps on the neighbor seam. Surely, the complication of the scheme in this case is impractical, because the neighbor stope, being quite far from the considered one, does not influence significantly on the considered area and it may be studied separately. We have mentioned about it just to underline that the comparison of observed and simulated spatial distributions serves for validation and improvement of geometrical features. It may also disclose the presence and influence of nearby faults.

Fig. 3 shows the spatial distribution of events in the projection to the vertical plane orthogonal to the stope face. The abscissa in this figure is in the direction opposite to the direction of the front advance (the mining front moves to the left). The origin coincides with the position of the stope at the beginning of the time step. It can be seen that the events tend to group behind the lines shown by dashes, which pass through the origin and comprise angles approximately $124^o$ with the advance direction. It appears that a considerable part of events (about 80%) is within the strips of width 20 m near these lines. Actually these are the strips in the roof and the floor, where the *stress concentration* ahead of the mining front is changed to *stress unloading* caused by extraction of a seam. This effect also appears in specific distributions of B, P and T directions presented below when discussing the mechanisms of events.

The distribution of Fig. 3 also shows that almost no events occur in the seam ahead of the mining front. It is explained by high concentration of compressive stresses which exclude both tensile and shear events. Clearly, the absence of events in this zone does not mean that it is safe as concerns with unfavorable effects of rock pressure. In fact, this zone is quite dangerous because of high amount of elastic energy it. This shows the need to combine the observations and modeling of seismicity with observations and modeling of geomechanical quantities like stresses.

*3.2. Frequency-magnitude and extreme distributions. Risk evaluation*

Scalar characteristics of the strength of an event (energy, value of the seismic potency and moment) are commonly presented in log-log coordinates by a dependence of Gutenberg-Richter type. To consistently compare simulated events with those observed, it is reasonable not to take into account simulated events with the energy below the threshold of measurements in a mine. In the considered example of mining in the Vorkuta coal district, the threshold was 1 J. Respectively, when presenting distributions of energy for simulated events, we account for merely events with the energy exceeding this lower limit.

Fig. 4 presents the dependencies of Gutenberg-Richter type for the simulated (dashed line) and observed (solid line) events of April 2013. The agreement of the curves is quite poor, because surprisingly small number of weak events (with the energy on the level 1 J) was registered in the mine. The agreement may be improved by excluding simulated events with the energy on this level. Still there are doubts about the quality of observations as concerns with events with so small energy. This implies that accounting for weak simulated events is impractical when their energy is well below a reliable level of energy for the weakest registered events. In such cases, it is reasonable to exclude these events from the analysis of frequency-magnitude dependencies.

Turn now to the risk analysis. According to the classical definition (see, e.g. [15, 16, 19]), the seismic hazard $R(W^*, T)$ is the probability that a seismic event not slighter than $W^*$ ($W$ is a considered measure of the event strength, say, energy) occurs in the time interval $T$. Under the assumption that the process generating seismic events is stationary, we have

$$R(W^*, T) = 1 - \exp[-\lambda T(1 - F(R(W^*)))]$$

where $\lambda$ is the mean activity rate of seismic activity (the parameter of the Poisson distribution), $F(x) = P(W < x)$ is the cumulative distribution function of $W$.

A proper assigning of $\lambda$ and $F(x)$ is crucial for evaluating the seismic hazard. The usual estimation of the parameter $\lambda$ is provided by the number of seismic events occurred per unit time. For *a 3-day step considered*, the total number of simulated events is 273. Hence, with this time scale, $\lambda = 273$. To obtain the function $F(x)$, Fig. 5 presents distributions of energy for the observed (Fig. 5a) and simulated (Fig. 5b) events. Step-functions correspond to empirical distributions, solid lines show their best-fit analytical approximations by the so-called Standard Generalized Extreme Value Distributions (SGEVD):

Fig. 5a,b

$$F(x) = \exp[-(1 + kz)^{\frac{1}{k}}],$$

where $z = (x - \mu)/\sigma$. For the observed events $k = 0.39$, $\sigma = 775$, $\mu = 1296$. For the simulated events, $k = 0.69$, $\sigma = 607$, $\mu = 257$. The SGEVD are of the Frechet type ($k > 0$). Again the agreement between the parameters corresponding to the observed and simulated events is not satisfactory. Obviously, it is explained by surprisingly small number of observed events with low energy. Thus we focus on the simulated events. We are interested in an approximation, which better describes their empirical distribution than SGEVD. It appears that the best-fit approximation is given by the Weibull distribution:

$$F(x) = 1 - \exp[-(x/\beta)^\alpha]$$

with $\alpha = 0,34$ and $\beta = 473$.

Numerical short-term analysis of several subsequent steps shows that the results, discussed above, are typical as concerns with both the number of seismic events and the energy distribution. The parameters of the best-fit Weibull distributions are within the ranges $0.29 < \alpha < 0.34$ for $\alpha$, and $433 < \beta < 665$ for $\beta$. Having the found Poisson parameter $\lambda$ and the cumulative distribution function $F(x)$, the risk evaluation is of immediate use.

*3.3. Distributions of geometrical parameters of the source mechanism*

The mechanism of a seismic event is characterized by the tensor of seismic moment (e.g. [11-14, 16]) or, alternatively, by the tensor of seismic potency [16]. When fixing a reference shear modulus, these tensors are practically equivalent differing only by the scale of the magnitudes. Then their geometrical characteristics are the same. Below being interested in geometrical parameters of a source, we shall not distinguish between the tensors.

Conventionally, the geometrical features of the mechanism are presented by stereographic projections of the nodal planes, the unit normal to the planes, the zero B direction, the pressure P and tension T directions. When considering statistical distributions of many events, it is inconvenient to use the projections of the planes themselves: there are too many *curves* covering the entire circle. Rather it is reasonable to use actually equivalent representation by *points* corresponding to the unit normal to the planes. Below when talking about the stereographic projection of a nodal or strain plane, we shall mean the unit normal to it. For certainty, the normal is defined in such a way that its angle with the upward vertical direction is within the common range $[0, \pi]$ of the dip angle. We associate the nodal plane I with the plane at which an event occurs; then the nodal plane II has the normal collinear to the vector of shear displacement on the flaw surface.

Fig. 6a,b

For a typical mining step, the distributions of nodal planes I and II are shown in Fig. 6a and b, respectively. The first of them does not look informative; there is no distinct grouping of points. The distribution of the nodal plane II appears a bit more instructive. We may see a trend to grouping of points along the North-South diameter. Still the trend is not quite pronounced.

Fig. 7a,b,c

In contrast, the distributions of zero B, P and T directions, presented in Fig. 7a, b and c, respectively, are notably more informative. From Fig. 7a, it follows that the zero B directions are grouped near the periphery of the East-West diameter. This means that the strain planes, defined by B, tend to group near the *vertical* plane with the strike along the North-South line. Note that this is the plane of Fig. 3.

The distributions of P and T directions, which are orthogonal to each other and located in the strain plane, further clarify the features of dominating mechanisms of simulated events. From Fig. 7b it can be deduced that P directions have the dip angle grouping near the origin what means that these directions are mostly horizontal. Notable part of them has the major component along the North-South line. Respectively, Fig. 7c shows that T directions are well apart from horizontal plane. Since the shear vector comprises $45^o$ with P and T directions, we conclude that notable part of shear displacements occurs in the direction comprising $135^o$ with the direction of the stope propagation.

Recall now that as follows from the distribution in Fig. 3, many events occur in a narrow strip, which corresponds to transition from stress concentration to unloading. This implies that grouping of distributions in Fig. 7 is in close connection with quite general changes in stresses induced by a mining step.

We conclude that the analysis of B, P and T directions, plotted in the stereographic projection, may reveal significant features of seismicity. They are closely connected with the changes of stresses. Hopefully, these distributions may serve for better interpretation of observed seismicity.

## 4. CONCLUSIONS

The conclusions of the papers are summarized as follows.

(i) The input data used in a basic code for modeling of seismicity, may be restricted with available data on major geometrical, geological and mechanical features of a region of interest. When playing scenarios, additional parameters not available from direct observations, such as cohesion and angle of friction at a fault, may be included into the input data of the basic code.

(ii) The minimal input data, used in subroutines complementing the basic code and defining the sources of events, include the tensile strength, cohesion and friction angle of a flaw and, if wanted, a desired portion of simulated aseismic events. The location, orientation and sizes of the initial flaw-sources may be either seeded randomly or found by checking the strength criteria at points of a prescribed mesh and distinguishing lobes where the criteria are met. The choice between the two approaches is made depending on a particular objective of modeling. The statistical approach, employing *a priori* information on the average energy of observed events, provides wider options for the analysis of the output data on distributions of frequency-magnitude type. The approach, based on checking the strength conditions, has the advantage of avoiding *a priory* information about the sizes of flaws. It also automatically focuses on areas, which are most prone to produce seismicity.

(iii) The output parameters of numerical simulation of seismicity may be allocated to three major groups of output distributions with distinct practical applications. The first group includes the data on temporal and spatial distributions. These distributions are of special value for improving the input data on geometrical and structural features of an area considered. The second group employs the data on the magnitude of events. It includes distributions of the frequency-magnitude type. These distributions are of exceptional need when evaluating the risk of dangerous strong events. The third group includes the data on distributions of geometrical parameters of the source mechanism (orientation of nodal planes, B, P and T directions). It is especially important when establishing and using the connection between the stresses and seismicity. It may serve for improving the input data on in-situ stresses and for better interpretation of observed seismicity.

**Acknowledgment.** The work was supported by the Russian Scientific Fund (Grant # 15-11-00017).

# List of figure captions.

Fig. 1. The plan view of the longwall opening 512-z V in the Severnaya mine in April 2013

Fig. 2. The distribution of simulated seismic events in the seam plane for 10 m advance of the stope

Fig. 3. The distribution of simulated seismic events in the vertical plane orthogonal to the stope

Fig. 4. Dependencies of Guttenberg-Richter type corresponding to the observed (solid line) and simulated (dashed line) seismicity

Fig. 5. Empirical (step-functions) and analytical (continuous solid lines) distributions of probability to have events with energy exceeding x:

a) distributions for observed events

b) distributions for simulated events

Fig. 6. Stereographic projections of the unit normal to

a) the nodal plane I

b) the nodal plane II

Fig. 7. Stereographic projections of the unit directions:  a) B (zero strain),  b) P (pressure), c) T (tension)

Figure 1.

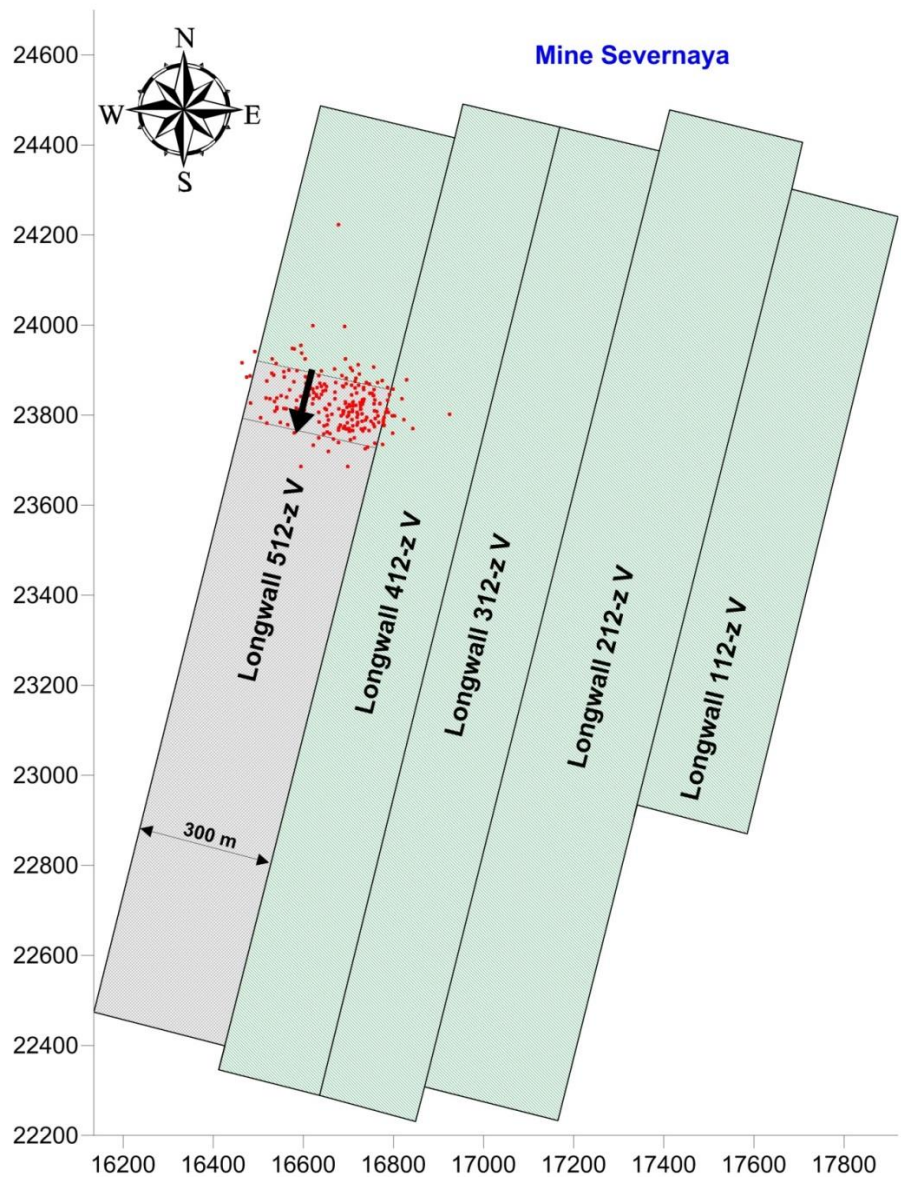

Figure.2

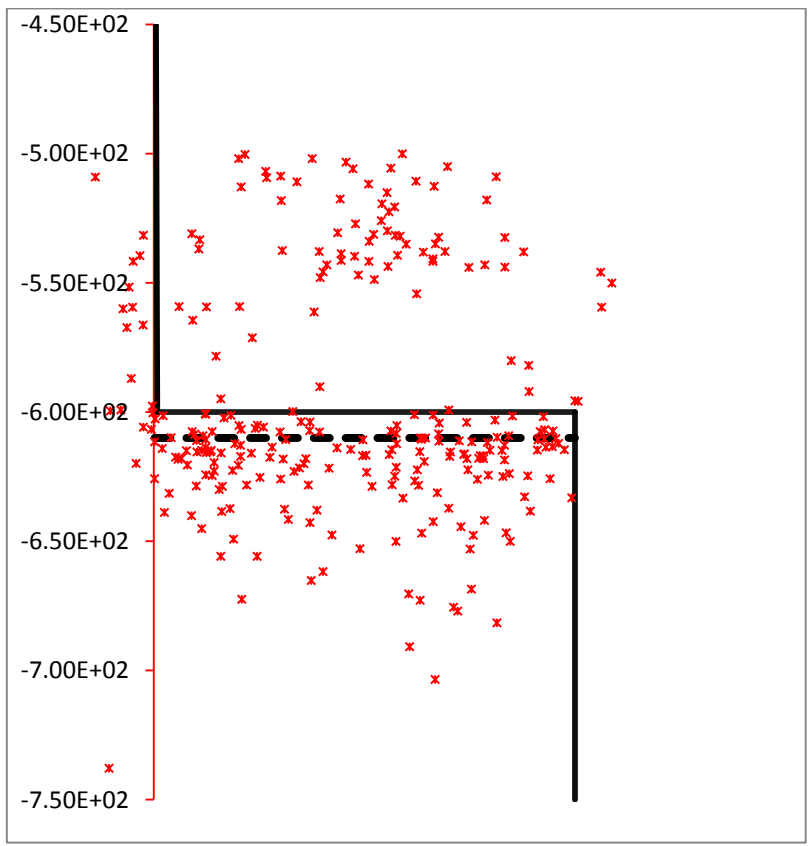

Figure. 3.

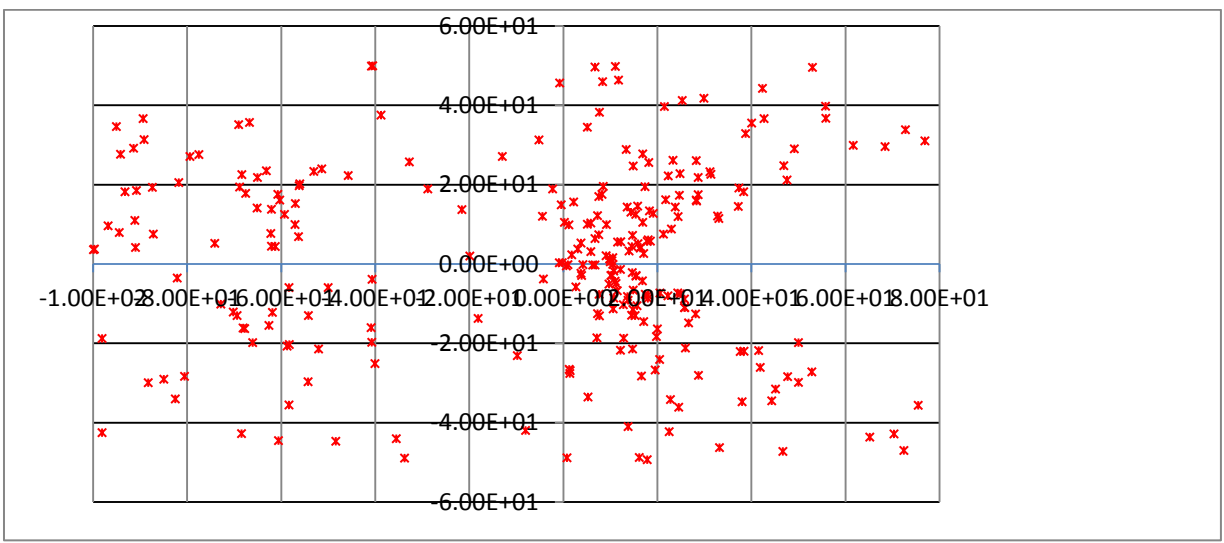

Figure. 4.

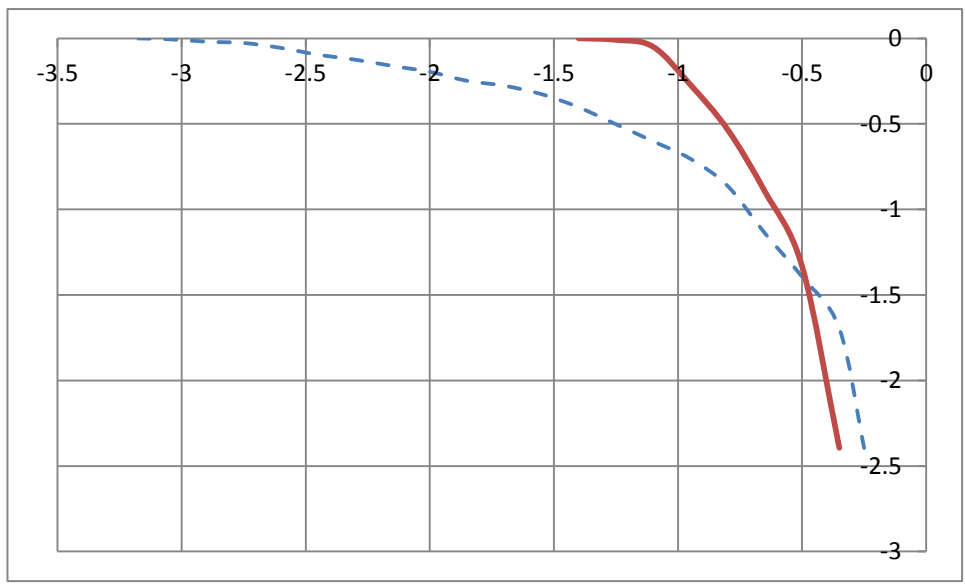

Figure 5. a, b

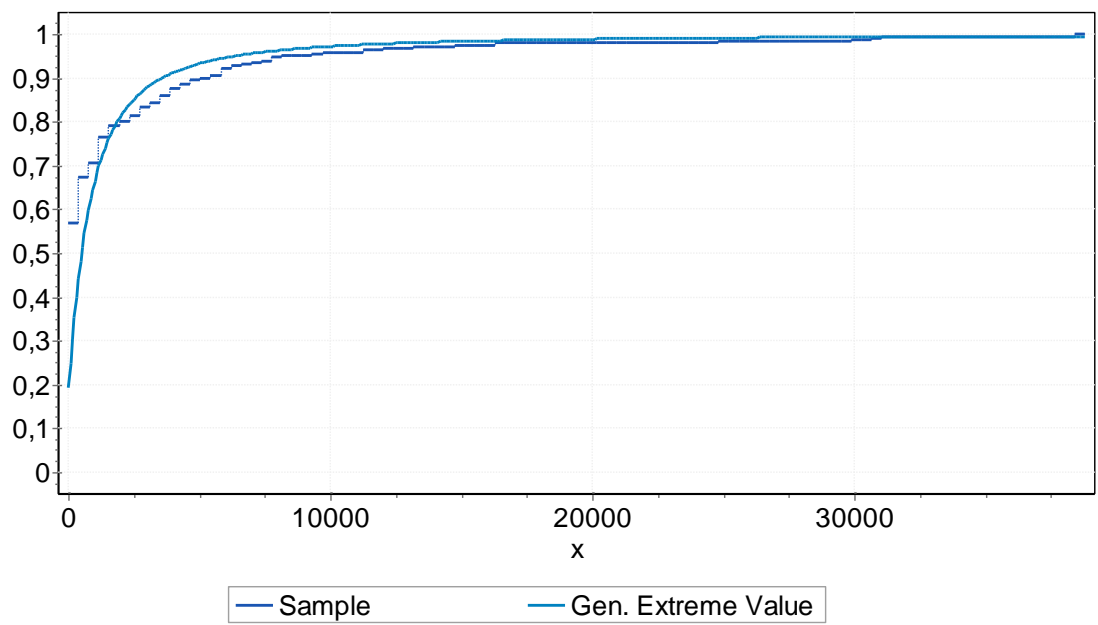

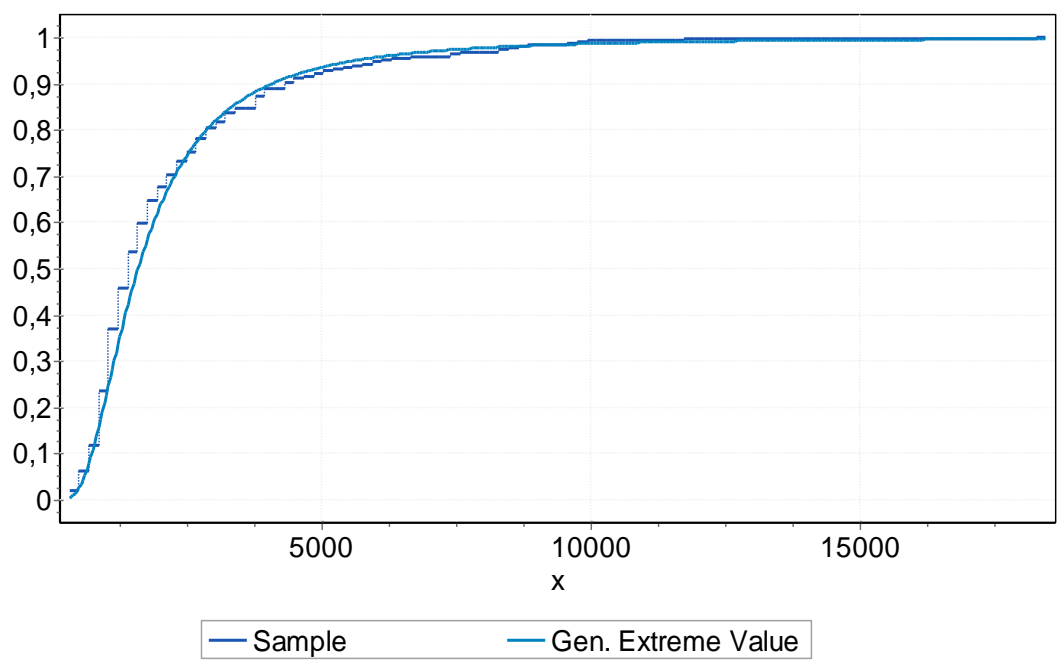

Figure 6.

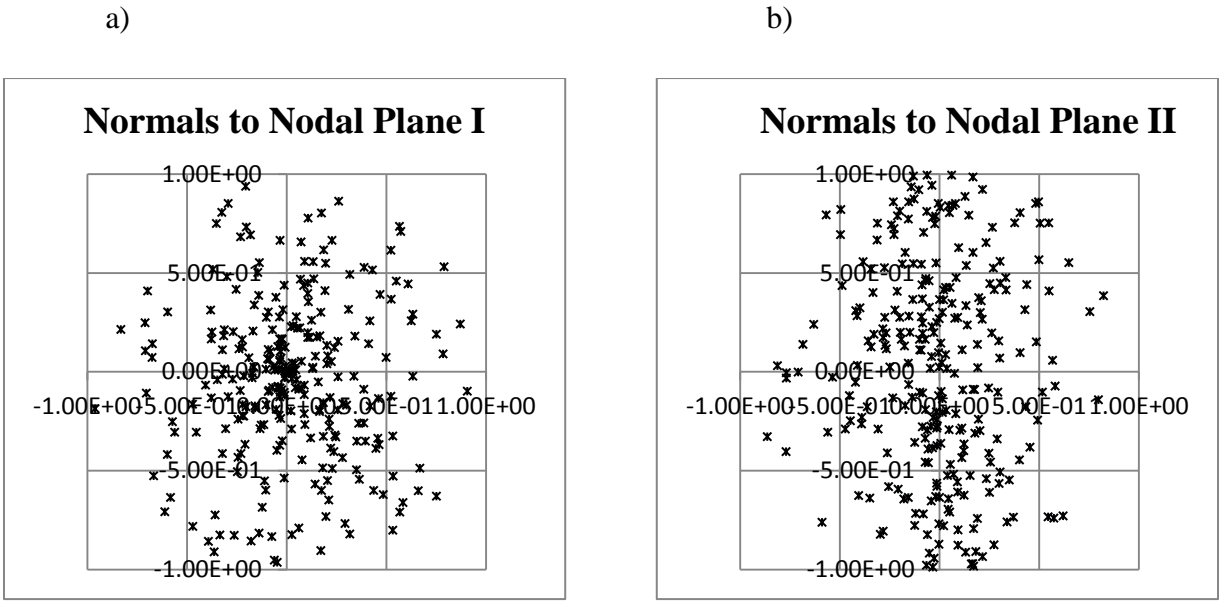

a) b)

Figure 7.

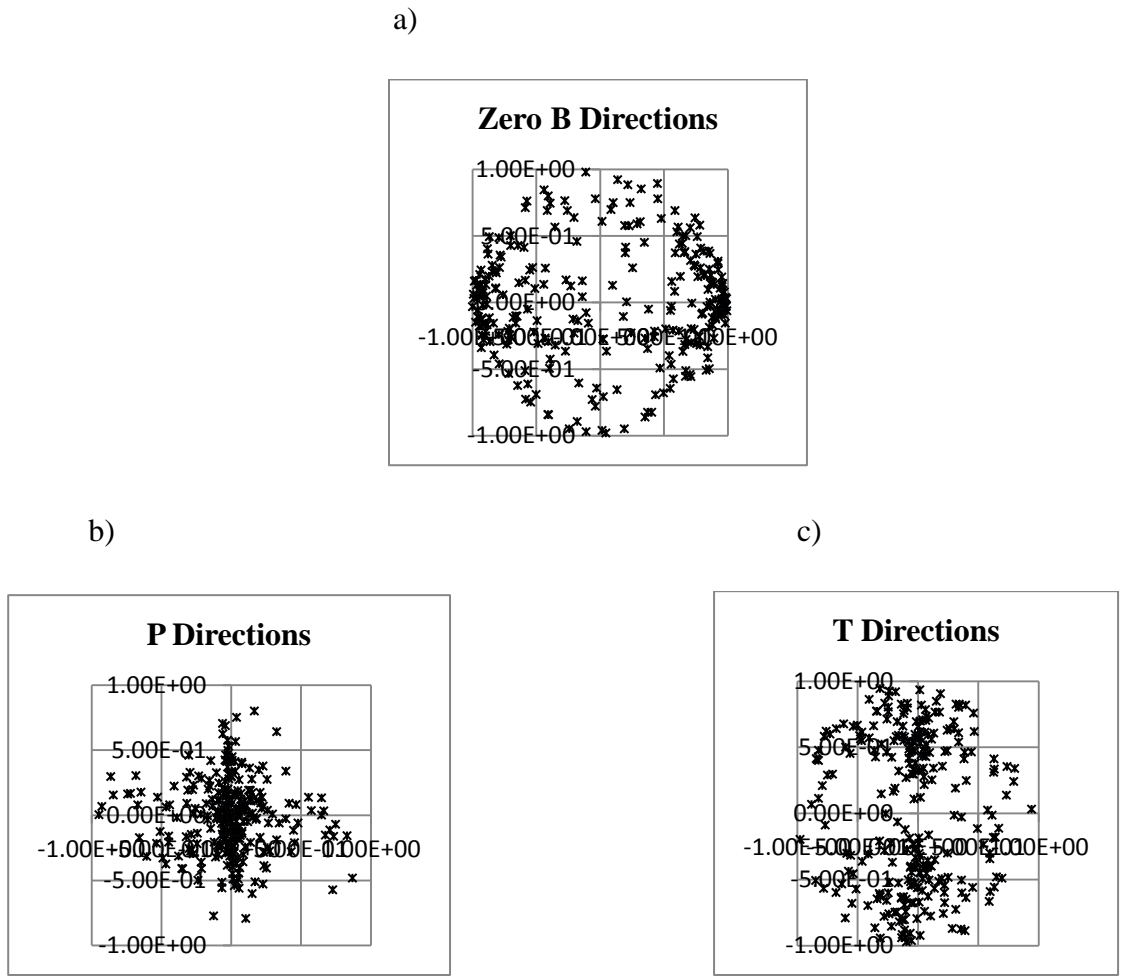

Fig. 7. Stereographic projections of the unit directions

a) B (zero strain)

b) P (pressure)

c) T (tension)